\documentclass[onecolumn,A4]{article}

\usepackage{cite}
\usepackage{anysize}

\usepackage{mathrsfs}
\usepackage{amsmath}
\usepackage{amssymb}
\usepackage{graphicx}
\usepackage{subfigure}
\usepackage{amssymb}
\usepackage{amsthm}
\pdfoutput=1
\usepackage{pgf}
\usepackage{pdfpages}
\DeclareGraphicsExtensions{.png}

\begin{document}

\vskip 1cm
\marginsize{3cm}{3cm}{3cm}{1cm}

\begin{center}
{\bf{\huge Signal Formation in THGEM-like Detectors}}\\
~\\
Purba Bhattacharya$^{a*1}$, Luca Moleri$^b$, Shikma Bressler$^a$\\
{\em $^a$ Department of Particle Physics and AstroPhysics, Weizmann Institute of Science, Herzl St. 234, Rehovot - 7610001, Israel}\\
{\em $^b$ Department of Physics, Technion - Israel Institute of Technology, Haifa - 3200003, Israel}\\
~\\
~\\
~\\
~\\
~\\
{\bf{\large Abstract}}
\end{center}

\noindent Numerical simulations were used to study signal formation in a Thick Gaseous Electron Multiplier (THGEM) and in THGEM -based Thick-WELL (THWELL) and Resistive-Plate WELL (RPWELL) detectors. The signal shapes were simulated in mixtures of Argon and Neon with $5\%$ Methane under irradiation with soft x-rays and muons. Anode-induced raw signals were convoluted with the response functions of charge-sensitive and current-sensitive pre-amplifiers. The simulation toolkit was validated by the good agreement reached between the simulated and measured response, with different pre-amplifiers. It indicates that our simulations framework provides valid insight into the inherent complex dynamical processes of the various detectors.

\vskip 1.5cm
\begin{flushleft}
\end{flushleft}

\vskip 1.5in
\noindent
{\bf ~$^*$Corresponding Author}: Purba Bhattacharya\\

E-mail: purba.bhattacharya85@gmail.com

\newpage


\section{Introduction}

Signals resulting from radiation-induced charges in gaseous radiation 
detectors, drifting under electric field and multiplied in a gas-avalanche 
mode - provide information on the deposited energy and position of the 
interacting event. 
The current induced on a specific electrode by a drifting charge can be 
calculated from the Shockley-Ramo Theorem \cite{paper1, paper2}. 
The induced charge in the detector is constant along its drift path, 
under an applied electric field. 
The shape of the currents induced by drifting electrons and ions, depends 
on the detector type and geometry, on the electric-field distribution and 
on the charge mobility in a given gas \cite{paper3, paper4, paper5, paper6}.

Charge-sensitive preamplifiers are often used to provide an output voltage 
pulse of amplitude proportional to the integral of the varying-in-time 
input current pulse. 
The output pulse can be obtained by convoluting the input current with the 
pre-amplifier's response (transfer function).
For more details see \cite{paper4, paper5}.

In this work, we have used Garfield \cite{paper7, paper8} in combination with 
neBEM \cite{paper9, paper10}, 
Magboltz \cite{paper11, paper12} and Heed \cite{paper13, paper14} software packages for 
evaluating the signal formation in some Thick Gaseous Electron Multiplier 
(THGEM)-like 
detectors \cite{paper15, paper16}: standard THGEM, Thick WELL (THWELL) \cite{paper17} and 
Resistive-Plate WELL (RPWELL) \cite{paper18}. 
Their schematic descriptions are shown in Figure \ref{Detector}. 
They share similar geometrical and material features.

A comprehensive comparison of their simulated signal shapes is presented 
in this work and compared to experimental results. 
Note that similar framework was used in the past for simulating the signal 
shapes in Micromegas \cite{paper19} and RPC \cite{paper20} detectors. 
In the process of evaluating the signal shapes from the THGEM-based detectors, 
we have also convoluted the raw signals with the transfer functions of the 
different pre-amplifiers employed to reproduce the experimental 
conditions.

The detector geometries are discussed in section 2. The simulation method 
is detailed in section 3. The experimental setup is described in section 4, 
followed by the results in section 5 and a discussion in section 6.

\section{THGEM-based Detectors}

\begin{figure}[hbt]
\centering
\subfigure[]
{\label{Detector-1}\includegraphics[height=2.5cm]{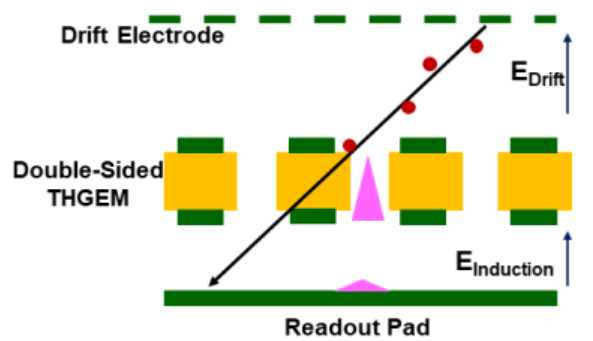}} \\
\subfigure[]
{\label{Detector-2}\includegraphics[height=2.5cm]{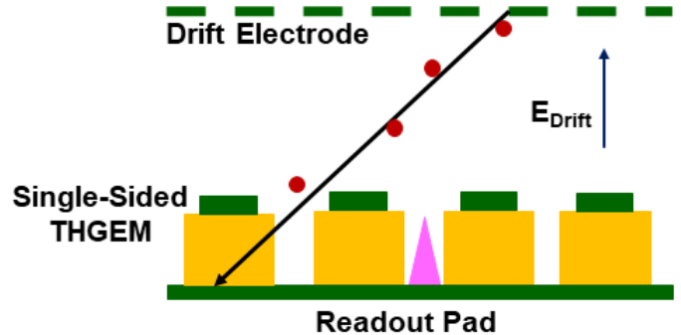}}
\subfigure[]
{\label{Detector-3}\includegraphics[height=2.7cm]{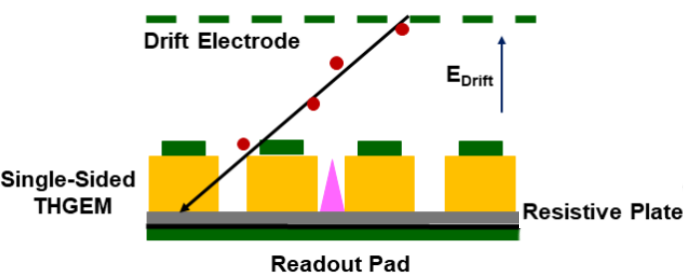}}
\caption{A schematic view of the (a) double-sided THGEM with an induction gap, (b) THWELL and (c) RPWELL detectors..}
\label{Detector}
\end{figure}

The THGEM \cite{paper15, paper16, paper21, paper22} is a simple and robust 
detector suitable for 
applications requiring large-area detection surfaces with moderate spatial 
resolution ($\sim0.2~-~0.3~\mathrm{mm}$ RMS). 
In its original configuration, a THGEM electrode is made of a 
$\mathrm{sub-mm}$ thick double-sided printed-circuit board (Cu-clad on 
both surfaces) having  
mechanically-drilled holes typically $\sim0.5~\mathrm{mm}$ in diameter with chemically etched rims, patterned 
over the surface with $\sim1~\mathrm{mm}$ pitch (Fig. 1a). 
Single-sided THGEM electrodes (copper-clad on its top side only) are used for WELL configurations, in which a 
readout anode is coupled directly (THWELL, Fig. 1b) or via resistive material 
(Resistive WELL \cite{paper17} and the RPWELL (Fig. 1c)) to the 
multiplier's bottom surface. 
The resistive coating of the anode of its coupling through a resistive plate 
aim at quenching occasional discharges. 
A stable operation of RPWELL detectors in muon- and hadron-beams, was 
demonstrated in neon- and argon-based gas mixtures \cite{paper23, paper24, paper25, paper26}.

In its standard configuration (Figure \ref{Detector-1}), the THGEM electrode is preceded 
by a conversion and drift gap. 
Radiation-induced ionization electrons drift into the THGEM holes, where
multiplication occurs under a high electric field. 
The resulting avalanche electrons are extracted into a few-mm wide 
induction gap and drift towards the readout anode, inducing a signal. 
As long as the field strength in the induction gap is below the multiplication 
threshold, the induced signal (measured on the readout anode) is characterized 
by a fast rise time (of a few $\mathrm{nsec}$). 
The anode is sensitive essentially only to the motion of avalanche electrons 
along the induction gap; the ion movement occurs within the THGEM hole, 
far from the anode, and has a negligible effect on the pulse shape \cite{paper27}. 
In the THWELL (Figure \ref{Detector-2}), higher gains could be obtained at a given applied 
voltage across the THGEM electrode, due to the larger electric field within 
the closed holes. 
The signal induced on the THWELL anode plate is characterized by a fast rise
(avalanche electrons) followed by a slow avalanche-ion component \cite{paper27}. 
In the RPWELL, the electrode is coupled to a thin plate of high bulk 
resistivity ($\sim10^{9}~-~10^{12}~\Omega\mathrm{cm}$), with the patterned 
readout anode placed behind \cite{paper18}. 
Despite of the additional Resistive Plate (RP), the pulse shape on the 
anode is similar to that of the THWELL \cite{paper18, paper27}.

\begin{table}[hbt]
\caption{The investigated-detector configurations and operation parameters.}\label{THGEMVoltage}
\begin{center}
\begin{tabular}{|c|c|c|c|}
\hline
\bf{Detector} & \bf{Thickness} & \bf{Gas Mixture} & \bf{Voltage [V]} \\
\hline
THGEM & $0.8~\mathrm{mm}$ & $\mathrm{Ar}/(5\%)\mathrm{CH_4}$ & 2000 \\
\hline
THGEM & $0.6~\mathrm{mm}$ & $\mathrm{Ar}/(5\%)\mathrm{CH_4}$ & 1600 \\
\hline
THWELL & $0.8~\mathrm{mm}$ & $\mathrm{Ar}/(5\%)\mathrm{CH_4}$ & 1700 \\
\hline
THWELL & $0.4~\mathrm{mm}$ & $\mathrm{Ar}/(5\%)\mathrm{CH_4}$ & 1200 \\
\hline
RPWELL & $0.8~\mathrm{mm}$ & $\mathrm{Ar}/(5\%)\mathrm{CH_4}$ & 1750 \\
\hline
RPWELL & $0.8~\mathrm{mm}$ & $\mathrm{Ne}/(5\%)\mathrm{CH_4}$ & 975 \\
\hline
RPWELL & $0.4~\mathrm{mm}$ & $\mathrm{Ar}/(5\%)\mathrm{CH_4}$ & 1250 \\
\hline
RPWELL & $0.4~\mathrm{mm}$ & $\mathrm{Ne}/(5\%)\mathrm{CH_4}$ & 850 \\
\hline
\end{tabular}
\end{center}
\end{table}

In the present work, three types of detectors were investigated: 1) $0.6$ 
and $0.8~\mathrm{mm}$ thick double-sided single THGEM detectors (with 
induction gap), 2) $0.4$ and $0.8~\mathrm{mm}$ thick THWELL detectors and 
3) $0.4$ and $0.8~\mathrm{mm}$ thick RPWELL detectors. 
In the RPWELL, the resistive plate (RP) consisted of a $0.4~\mathrm{mm}$ 
thick Semitron ESD225 static dissipative acetal material with a bulk 
resistivity of $\sim10^{9}~\Omega·\mathrm{cm}$ \cite{paper28}. 
In all of the cases, the hole diameter and pitch are $0.5~\mathrm{mm}$ and 
$1~\mathrm{mm}$, respectively.
Avalanche electrons in the RPWELL holes, traversing the RP are evacuated 
to ground through the RP bottom surface - coated with a thin layer of 
graphite (surface resistivity $\sim3~\mathrm{M}\Omega/\mathrm{sq}$) \cite{paper26}. 
The geometrical and voltage configurations of the investigated detectors 
are summarized in Table 1. 
The drift gap in all configurations was kept at $5~\mathrm{mm}$, with 
spacers between the cathode mesh and the multiplier's top 
electrode; the drift field in all experiments was $0.5~\mathrm{kV/cm}$. 
The THGEM detector was operated with a $2~\mathrm{mm}$ wide induction gap 
(Figure \ref{Detector-1}), under an induction field of $2~\mathrm{kV/cm}$ (below charge 
multiplication onset).

\section{Numerical Model}

Garfield simulation framework \cite{paper7} has been used in the present work. 
The 3D electrostatic field simulations have been carried out with the 
neBEM (nearly exact Boundary Element Method) toolkit \cite{paper9}. 
HEED \cite{paper13} has been used for primary-ionization calculations and 
Magboltz \cite{paper11} - for computing drift, diffusion as well as Townsend and 
attachment coefficients.

The design parameters of the THGEM-based detectors considered in the 
numerical simulations are those described in section 2. The RP of the 
RPWELL detector was considered to be a dielectric - transparent to the 
signals induced on the anode \cite{paper29}. 
The basic electrode's cell structure has been repeated along 
both X- and Y-axes, to represent a real detector. 
The field configurations of the different detectors have been simulated 
using the voltage settings mentioned in Table 1.

Primary-electron clusters were generated for 5.9 keV photons and 
relativistic muons interacting with the detector medium; the latter being 
distributed along a track portion depositing energy within the drift gap 
(Fig. 2). 
The primary-electrons' drift process towards the THGEM 
electrode was simulated using the Microscopic tracking routine \cite{paper7}. 
In this procedure, a typical drift path proceeds through millions of 
collisions, each being classified as elastic or inelastic (including 
excitation, ionization, attachment etc.). 
The electrons are focused into the holes by the drift- and the 
hole's dipole fields, inducing avalanches within the holes. 
In this process, an electron starting from a given point is subject to 
collisions with gas molecules; at each step, a number of secondary electrons 
are produced according to the local Townsend and attachment coefficients; 
the newly produced electrons are traced like the initial ones. 
In parallel, the ions' drift lines are traced.

\begin{figure}[hbt]
\centering
\subfigure[]
{\label{Transfer-1}\includegraphics[height=4cm]{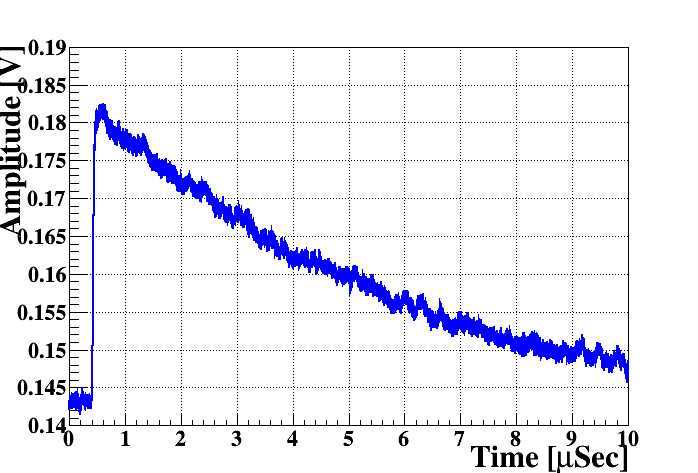}}
\subfigure[]
{\label{Transfer-2}\includegraphics[height=4cm]{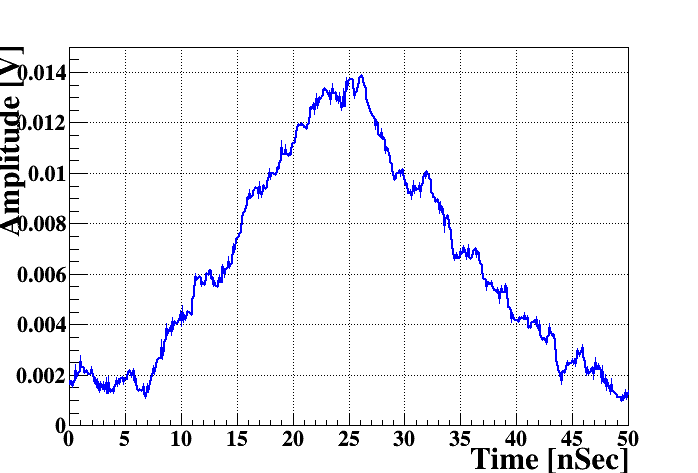}}
\subfigure[]
{\label{Transfer-3}\includegraphics[height=4cm]{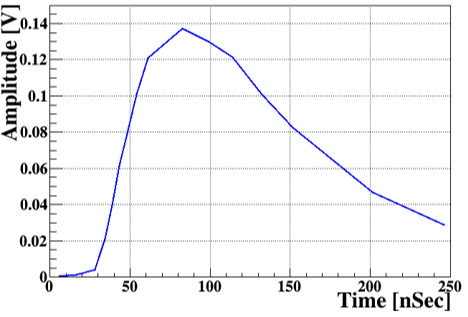}}
\caption{The measured response functions to a square pulse (with $200~\mu\mathrm{sec}$ width, $50~\mathrm{mV}$ amplitude), injected through an ORTEC pulse generator charge terminator ($2~\mathrm{pF}$, $100~\Omega$), of (a) charge-sensitive Canberra 2006 and (b) custom -made, fast current pre-amplifier. (c) The response function of the APV25 chip as reported in \cite{paper30}}
\label{TransferFunction}
\end{figure}

In a medium with perfect conductors and insulators, the current induced by 
a moving charge $\mathrm{q}$ onto an electrode can be calculated by means of the 
Shockley-Ramo theorem \cite{paper2, paper3}. The current $\mathscr{I}$ that flows into 
one particular electrode $i$ under the influence of a charge $q$ moving 
with velocity $v$ can be calculated using the following equation:

\begin{eqnarray}
\mathscr{I} = -q{\vec v \cdot \vec E_w \over V}
\end{eqnarray}

\noindent Here $E_w$ is the field created by raising this electrode to a potential 
$V$ and grounding all other electrodes, in the absence of charge.

The total charge $Q$ induced on the electrode is given by

\begin{eqnarray}
Q= \int_0^{\Delta t} i(t)dt = q \Delta V_w
\end{eqnarray}

\noindent $\Delta V_w$ is the weighting potential difference across which the charge has drifted.

We used the Shockley-Ramo theorem to calculate the induced current on a 
uniform readout anode. 
The current pulse was then convoluted with the readout electronics 
response function.
The output voltage from the preamplifier is then:

\begin{eqnarray}
V(t) = i(t) \otimes h(t)
\end{eqnarray}

The response functions of a charge sensitive and a fast pre-amplifier 
(used here) are shown in Figure \ref{TransferFunction}. 
The response function of the APV25 chip \cite{paper30} (Figure \ref{Transfer-3}) was taken from \cite{paper30}.

\section{Experimental setup}

\begin{figure}[hbt]
\centering
\includegraphics[height=4cm]{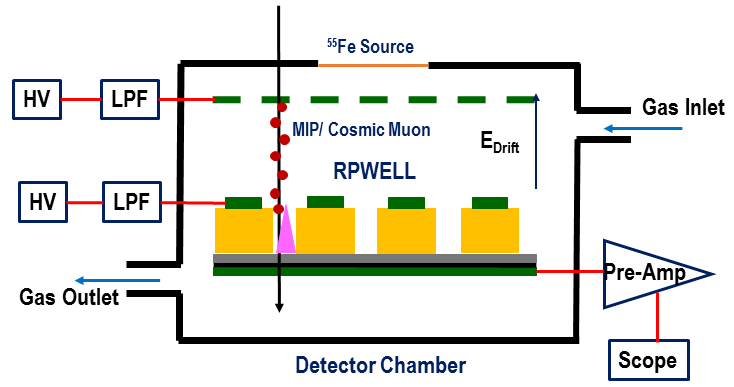}
\caption{A schematic view of the experimental setup. The detectors (shown is a RPWELL), placed in a vessel under gas flow, are irradiated by an external radiation source or charged particles. Signals from the anode (in a THGEM, after an induction gap; see text) are recorded with the pre-amplifier and processed by a digital oscilloscope. The detector electrodes are polarized through low-pass filters by high voltage power supplies .}
\label{ExptSetup}
\end{figure}

The experimental setup in lab, is shown in Figure \ref{ExptSetup}. 
The investigated detectors were placed in an aluminum vessel flushed with 
$\mathrm{Ar}/5\%\mathrm{CH_4}$ and $\mathrm{Ne}/5\%\mathrm{CH_4}$ at a flow of 
25 sccm (standard cubic centimeters per minute) at atmospheric pressure. 
The detectors were investigated at the Laboratory with 
$^{55}\mathrm{Fe}$-source 5.9 KeV x-rays and cosmic muons; complementary 
measurements with an RPWELL detector were carried out at CERN-SPS with 
150 GeV muons.

The drift electrode and the multiplier’s top electrode (and bottom electrode 
in the case of the THGEM) were polarized, through low-pass filters, by 
CAEN N1471H HV power supplies. 
In all configurations, the anode was at ground potential.
In the laboratory studies, anode signals (in the THGEM configuration 
readout-anode signals after the induction gap) were recorded through a 
charge-sensitive pre-amplifier (Canberra 2006; decay-time of 
$50~\mu\mathrm{sec}$) and a custom-made fast current pre-amplifier 
having shaping time of $\sim25~\mathrm{nsec}$. 
The preamplifier signals were processed by an Agilent Technologies 
InfiniiVision DSO-X 3034A digital oscilloscope. 
In all detector configurations, the voltage values across the multiplier 
(Table 1) were adjusted to keep similar gas gains of $\sim4000$.

In the CERN-SPS (H2 RD51 Test Beam area) accelerator measurements, the 
signal shapes of a $0.8~\mathrm{mm}$ thick RPWELL detector were studied in 
$\mathrm{Ne}/5\%\mathrm{CH_4}$. 
The signal pulses were recorded by an APV25 fast current pre-amplifier chip 
\cite{paper30} (shaping time of $75~\mathrm{nsec}$); they were further
processed and digitized by an SRS readout system \cite{paper31}. 
The latter has been frequently used to characterize the performance of 
MPGD detectors \cite{paper32, paper33}, including the THGEM and RPWELL ones 
\cite{paper23, paper24}.

\section{Results}
\subsection{Response to 5.9 keV photons}
\subsubsection{Simulated Raw Current Signals}

\begin{figure}[h]
\centering
\subfigure[]
{\label{Raw-1}\includegraphics[height=4cm]{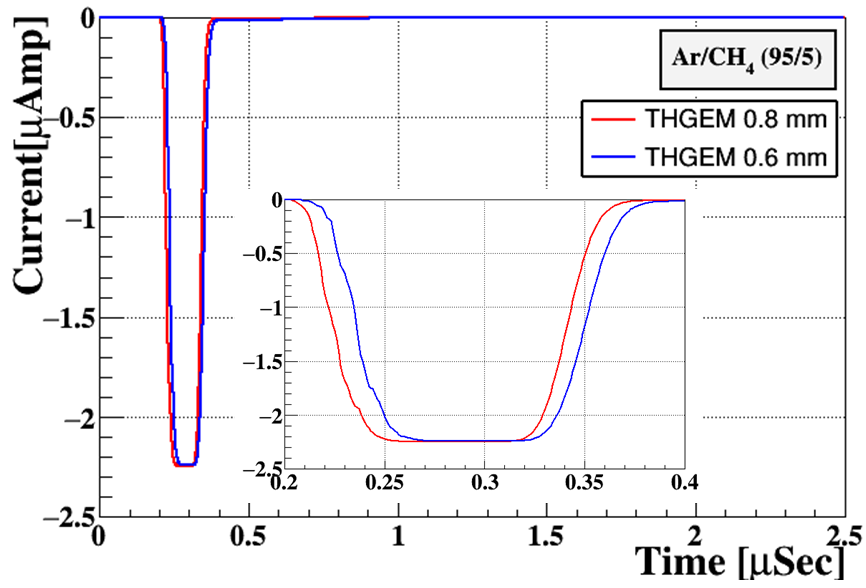}}
\subfigure[]
{\label{Raw-2}\includegraphics[height=4cm]{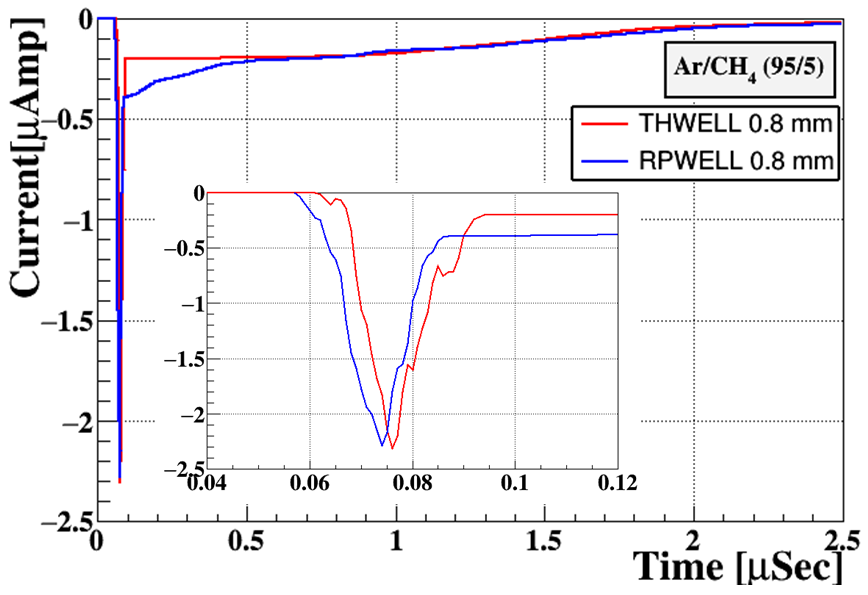}}
\subfigure[]
{\label{Raw-3}\includegraphics[height=4cm]{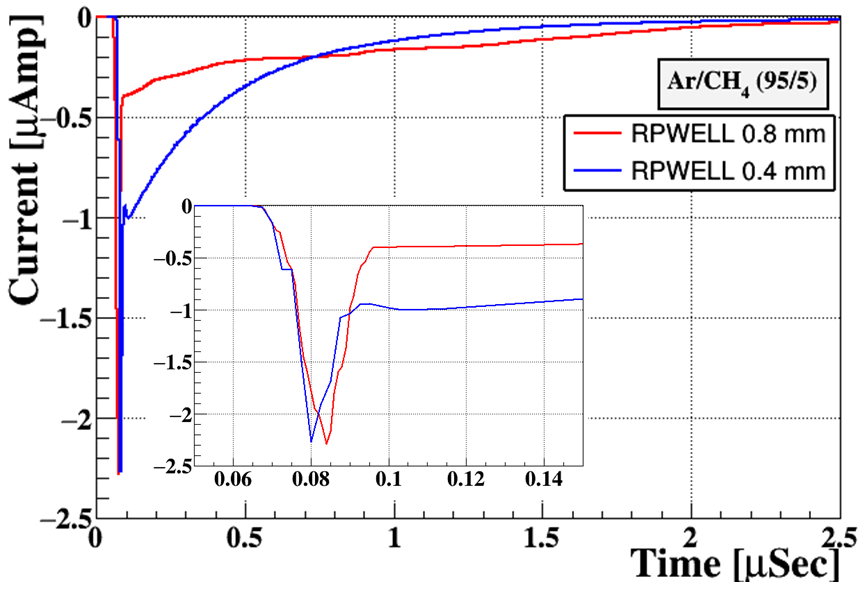}}
\subfigure[]
{\label{Raw-4}\includegraphics[height=4cm]{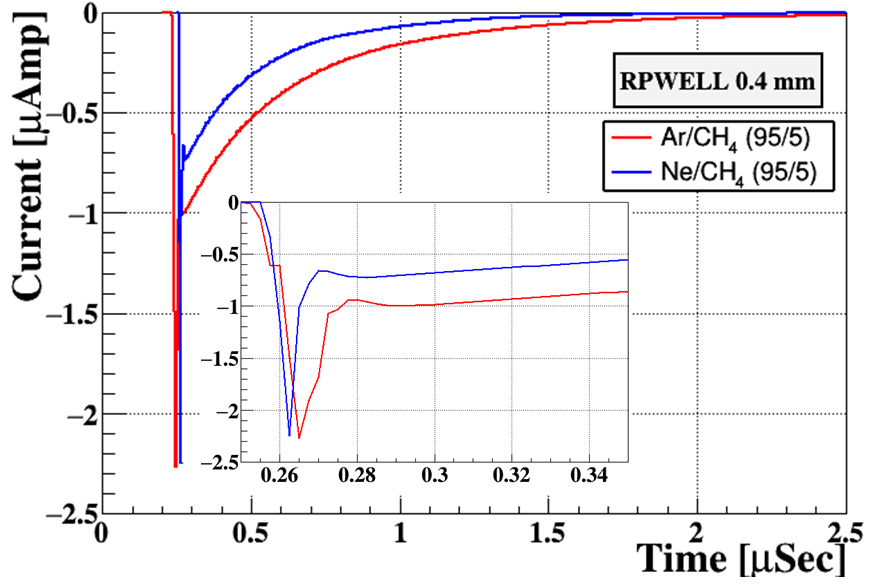}}
\caption{Simulated raw signals induced by 5.9 keV photons in (a) a THGEM with induction gap, (b) in $0.8~\mathrm{mm}$ thick THWELL and RPWELL detectors, (c) in $0.8$ and $0.4~\mathrm{mm}$ thick RPWELL detectors in $\mathrm{Ar}/5\%\mathrm{CH_4}$. (d) Signals in a $0.4~\mathrm{mm}$ thick RPWELL detector in $\mathrm{Ar}/5\%\mathrm{CH_4}$ and $\mathrm{Ne}/5\%\mathrm{CH_4}$. The fast signals are shown in the inset. The time axis of the insets is also in $\mu\mathrm{sec}$.}
\label{Raw}
\end{figure}

Typical simulated $5.9~\mathrm{keV}$ x-ray-induced raw current signals in 
the THGEM detector (Figure \ref{Detector-1}) are shown in Figure \ref{Raw-1} for $0.6$ and $0.8~\mathrm{mm}$ thick electrodes. 
The induced currents are due to the movement of avalanche electrons 
extracted from the hole, along the induction gap - towards the readout-anode 
plate. In this configuration, there are no ions drifting along the induction 
gap (no slow ion component). 
The signal length dictated by electron drift along the $2~\mathrm{mm}$ 
induction gap is $\sim120~\mathrm{nsec}$; the $0.2~\mathrm{mm}$ thickness 
variation does not affect this time by more than $\sim1\%$.

The raw current signals of the THWELL and RPWELL detectors are shown in 
Figure \ref{Raw-2}; the fast pulse is due to the avalanche electrons, whereas the 
long tail is the result of the ions' drift away from the 
avalanche head. 
The RPWELL current signals are peaked similarly to that of the
THWELL; however, the ion-component shape is different in both detectors. 
Due to the presence of the resistive layer (Figure \ref{Detector-3}), the weighting field 
in the RPWELL is different from that of the THWELL detector. 
But, in both cases, the ion tail reaches zero value within 
$\sim2~\mu\mathrm{sec}$. 
The raw current signal of RPWELL detectors having thickness of 
$0.8~\mathrm{mm}$ and $0.4~\mathrm{mm}$ are shown in Figure \ref{Raw-3}. 
The rise-time for the $0.8~\mathrm{mm}$ thick electrode is 
$\sim15~\mathrm{nsec}$, whereas for $0.4~\mathrm{mm}$ thick one it is 
$\sim11~\mathrm{nsec}$. 
Throughout this paper, the time required for the response to rise
from $10\%$ to $90\%$ of its maximum value is defined as the rise time of 
the signal . 
The decay time of the ion tail for the thicker electrode is 
$\sim1.7~\mu\mathrm{sec}$ in comparison to $\sim0.8~\mu\mathrm{sec}$ for 
the thinner one, consistent with the shorter ion drift time to the top 
electrode. 
A comparison of the raw signals of a $0.4~\mathrm{mm}$ thick RPWELL detector 
in $\mathrm{Ar}/5\%\mathrm{CH_4}$ and $\mathrm{Ar}/5\%\mathrm{CH_4}$ is 
shown in Figure \ref{Raw-4}. 
The ion tail in the Ne-mixture decays faster than that in the Ar-mixture; 
it is consistent with the faster ion velocity in the former.

\subsubsection{Charge-sensitive Pre-amplifier response: Measurements and Simulations}

\begin{figure}[h]
\centering
\subfigure[]
{\label{THGEMCharge-1}\includegraphics[height=3.5cm]{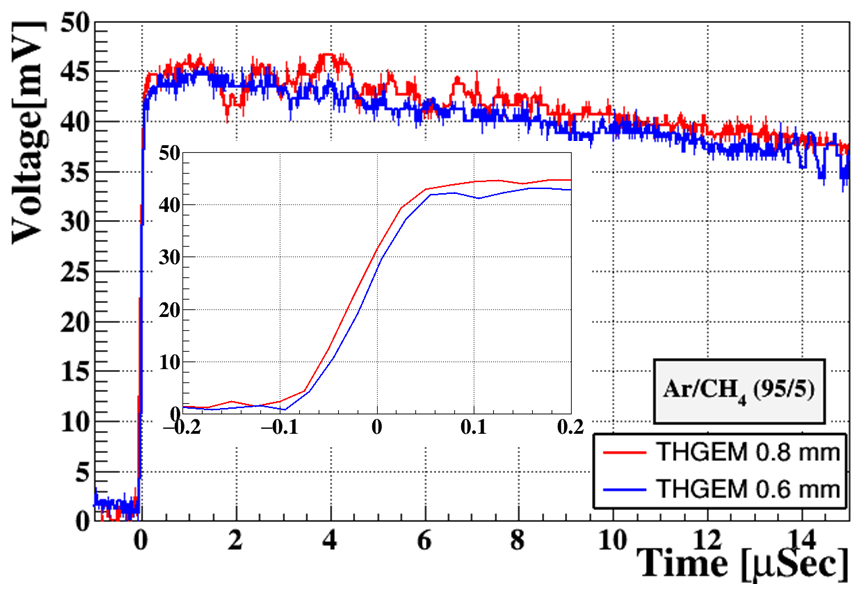}}
\subfigure[]
{\label{THGEMCharge-2}\includegraphics[height=3.5cm]{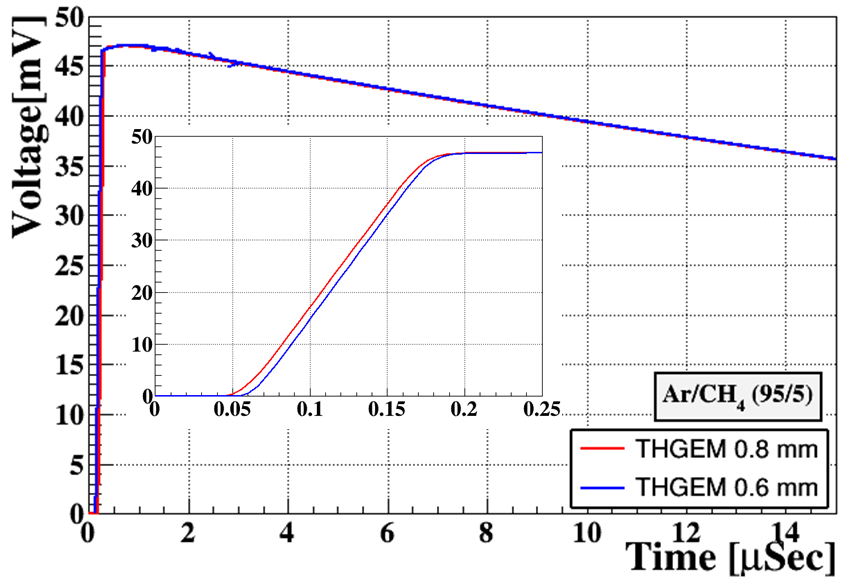}}
\caption{Measured and simulated charge-pre-amplifier pulses, induced by 5.9 keV photons in THGEM (a) measurement, (b) simulation. Electrode thickness: $0.8~\mathrm{mm}$ and $0.6~\mathrm{mm}$; gas: $\mathrm{Ar}/5\%\mathrm{CH_{4}}$. The pulse rise-times of the THGEM are shown in the inset. The time axis of the insets is also in $\mu\mathrm{sec}$.}
\label{THGEMCharge}
\end{figure}

\begin{figure}[h]
\centering
\subfigure[]
{\label{DetectorCharge-1}\includegraphics[height=3.5cm]{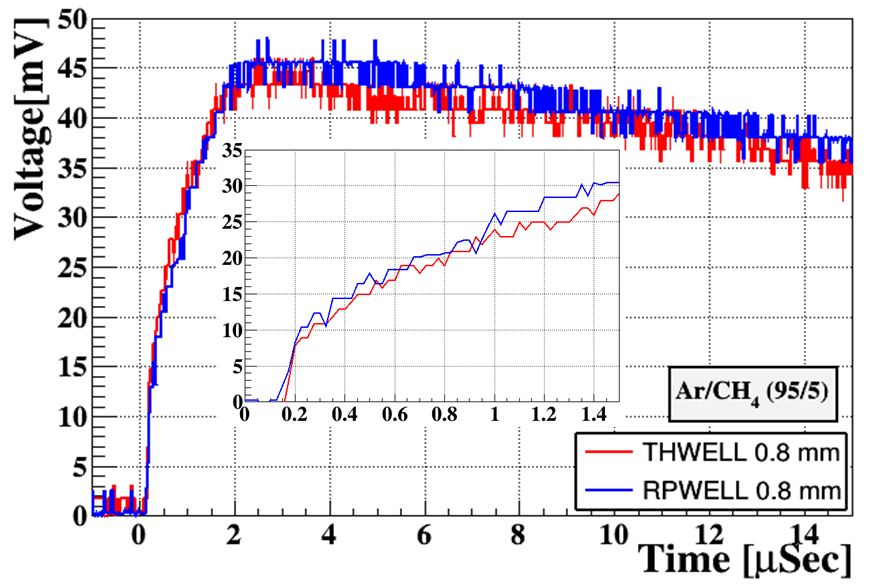}}
\subfigure[]
{\label{DetectorCharge-2}\includegraphics[height=3.5cm]{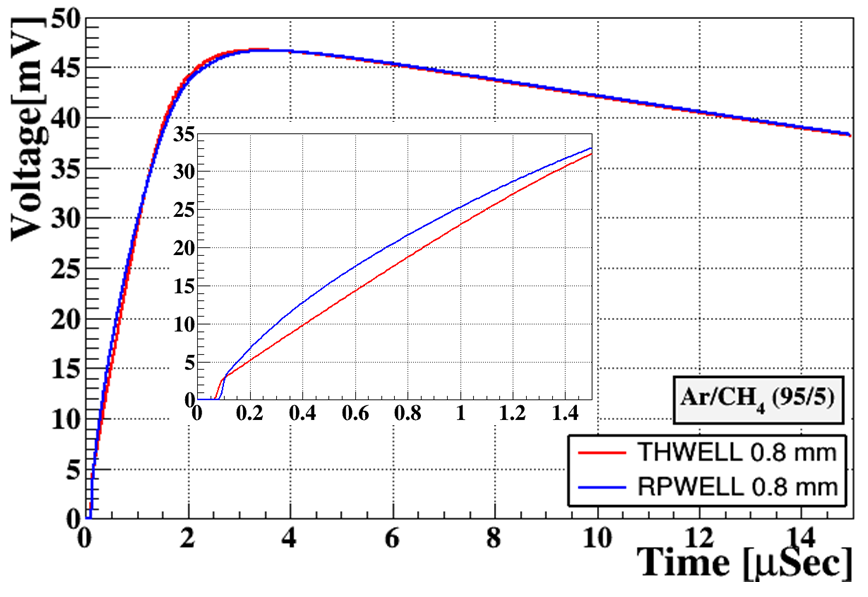}}
\caption{A comparison of the (a) measured and (b) simulated charge-signal shapes in $0.8~\mathrm{mm}$ thick THWELL and RPWELL detectors, operated in $\mathrm{Ar}/5\%\mathrm{CH_{4}}$. The fast and slow components of the two detectors are shown in the inset. The time axis of the insets is also in $\mu\mathrm{sec}$.}
\label{DetectorCharge}
\end{figure}

Charge-sensitive pre-amplifier signals measured on the anodes of the THGEM 
are shown in Figures 5(a). 
The rise-time of the THGEM pulse, reflecting the avalanche-electrons drift
to the anode, is $\sim100~\mathrm{nsec}$. 
The respective simulated charge signal is shown in Figures 5(b).
As discussed in Section 3, the raw current signals (Figure 4) are 
convoluted with the response function of the pre-amplifier. 
The decay time, determined by the decay constant of the pre-amplifier, is long, of the order of few tenths of $\mu\mathrm{sec}$. 
A comparison of the measured and simulated charge-signal shapes of THWELL 
and RPWELL detectors are shown in Figure 6. 
There is no significant difference in the signal shapes. 
The RPWELL detectors pulses have $\sim1.9$ and $\sim1.0~\mu\mathrm{sec}$ 
rise for respective electrode thicknesses of $0.8$ and $0.4~\mathrm{mm}$ 
(longer ion drift time along the holes) as shown in Figure 7. 
In all the above cases, the simulated data are in good agreement with the 
experimental data.

\begin{figure}[h]
\centering
\subfigure[]
{\label{RPWELLCharge-1}\includegraphics[height=3.5cm]{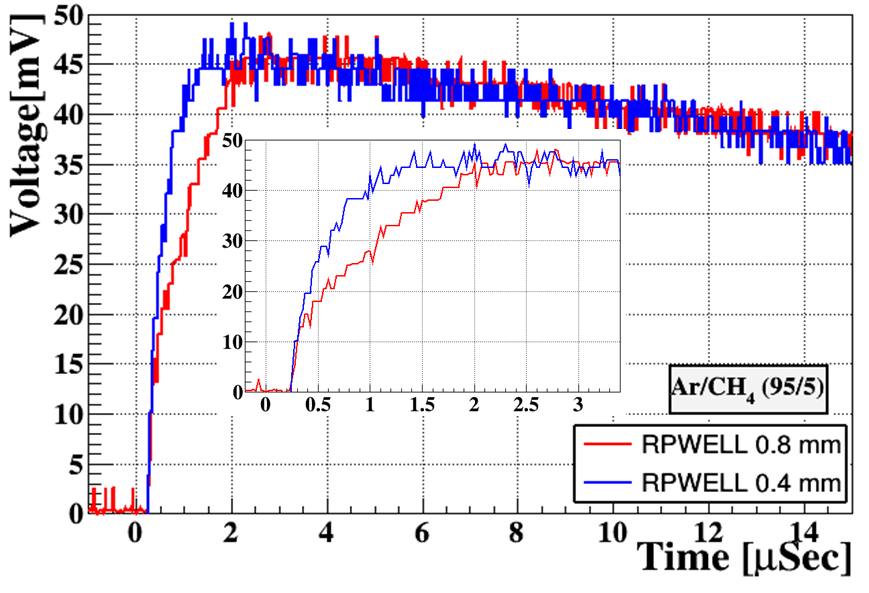}}
\subfigure[]
{\label{RPWELLCharge-2}\includegraphics[height=3.5cm]{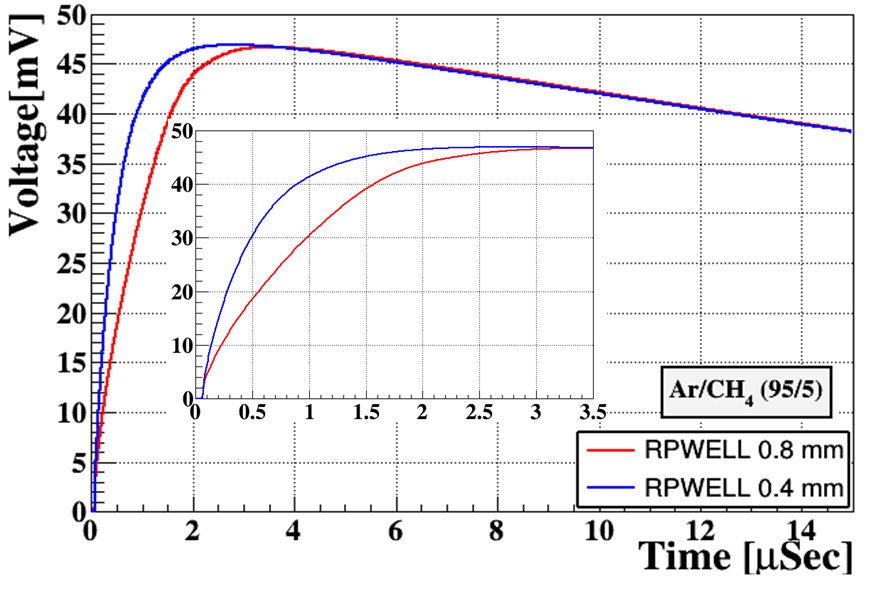}}
\caption{Measured and simulated charge-pre-amplifier pulses, induced by 5.9 keV photons in RPWELL (a) measurement, (b) simulation. Electrode thickness: $0.8~\mathrm{mm}$ and $0.4~\mathrm{mm}$; gas: $\mathrm{Ar}/5\%\mathrm{CH_{4}}$. The pulse rise-times of the RPWELL are shown in the inset. The time axis of the insets is also in $\mu\mathrm{sec}$.}
\label{RPWELLCharge}
\end{figure}

\begin{figure}[h]
\centering
\subfigure[]
{\label{GasCharge-1}\includegraphics[height=3.5cm]{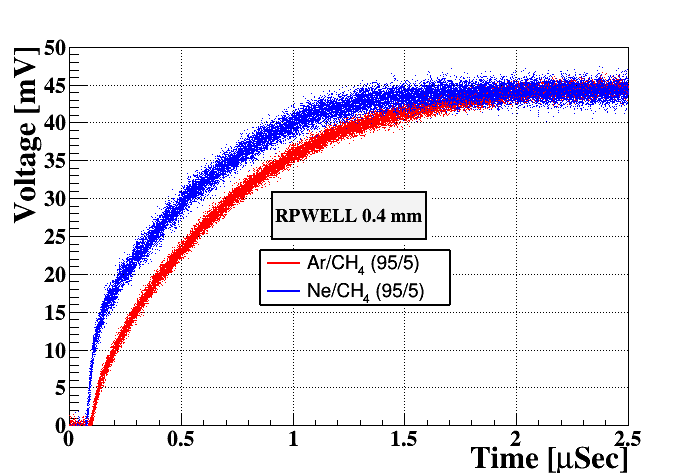}}
\subfigure[]
{\label{GasCharge-2}\includegraphics[height=3.5cm]{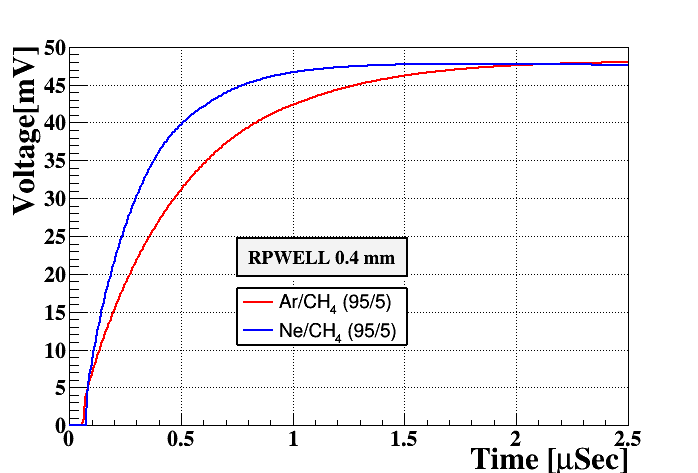}}
\caption{A comparison of the (a) measured and (b) simulated charge-signal shapes in a $0.4~\mathrm{mm}$ thick RPWELL detector, operated in $\mathrm{Ar}/5\%\mathrm{CH_{4}}$ and in $\mathrm{Ne}/5\%\mathrm{CH_{4}}$}
\label{GasCharge}
\end{figure}

A comparison of the measured and simulated signal shapes of the 
$0.4~\mathrm{mm}$ thick RPWELL detector in $\mathrm{Ar}/5\%\mathrm{CH_4}$ 
and in $\mathrm{Ne}/5\%\mathrm{CH_4}$ are shown in Figure \ref{GasCharge}. 
The simulated ion-component rise-time in the Ne mixture is faster than 
that in the Ar one, which is consistent with the fast ion mobility in 
Ne-based gas mixture. 
The experimental signals (\ref{GasCharge-1}) are in good agreement with the simulated ones (\ref{GasCharge-2}).

\subsubsection{Current Pre-amplifier response: Measurements and Simulations}

\begin{figure}[h]
\centering
\subfigure[]
{\label{THGEMCurrent-1}\includegraphics[height=3.5cm]{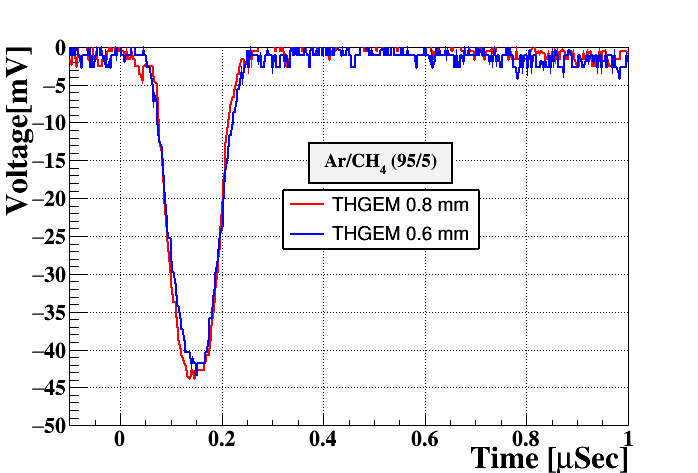}}
\subfigure[]
{\label{THGEMCurrent-2}\includegraphics[height=3.5cm]{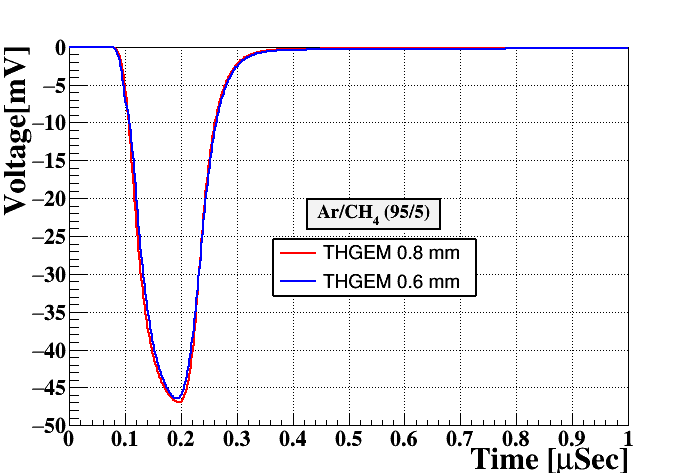}}
\caption{X-ray induced signals of $0.6~\mathrm{mm}$ and $0.8~\mathrm{mm}$ thick THGEM electrodes in $\mathrm{Ar}/5\%\mathrm{CH_{4}}$ gas mixture measured with a current-sensitive pre-amplifier (custom-made): (a) measurement, (b) simulation}
\label{THGEMCurrent}
\end{figure}

\begin{figure}[h]
\centering
\subfigure[]
{\label{DetectorCurrent-1}\includegraphics[height=3.5cm]{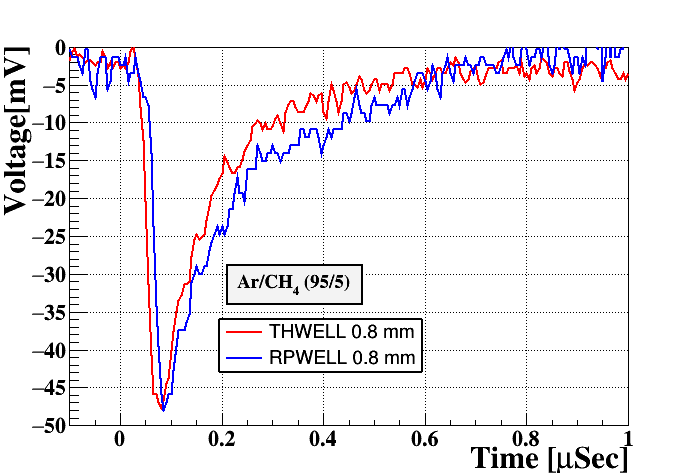}}
\subfigure[]
{\label{DetectorCurrent-2}\includegraphics[height=3.5cm]{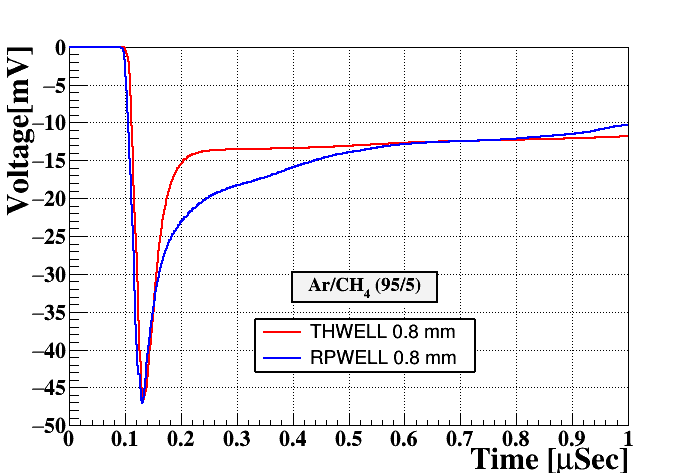}}
\caption{A comparison of the fast current signals in $0.8~\mathrm{mm}$ thick THWELL and RPWELL detectors in $\mathrm{Ar}/5\%\mathrm{CH_{4}}$ gas mixture measured from a custom-made current-sensitive pre-amplifier: (a) measurement, (b) simulation}
\label{DetectorCurrent}
\end{figure}

\begin{figure}[h]
\centering
\subfigure[]
{\label{RPWELLCurrent-1}\includegraphics[height=3.5cm]{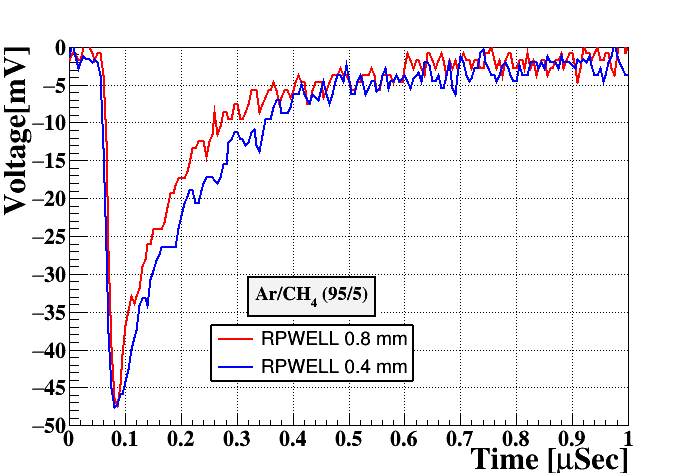}}
\subfigure[]
{\label{RPWELLCurrent-2}\includegraphics[height=3.5cm]{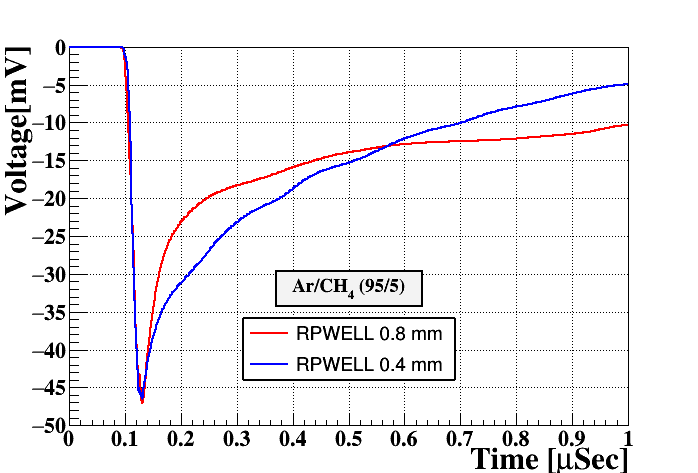}}
\caption{X-ray induced signals of $0.4~\mathrm{mm}$ an $0.8~\mathrm{mm}$ thick RPWELL detectors in $\mathrm{Ar}/5\%\mathrm{CH_{4}}$ gas mixture measured with a current-sensitive pre-amplifier (custom-made): (a) measurement, (b) simulation.}
\label{RPWELLCurrent}
\end{figure}

\begin{figure}[h]
\centering
\subfigure[]
{\label{GasCurrent-1}\includegraphics[height=3.5cm]{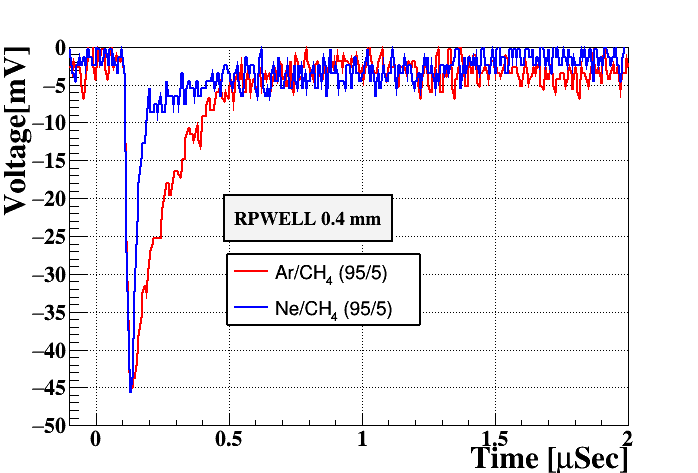}}
\subfigure[]
{\label{GasCurrent-2}\includegraphics[height=3.5cm]{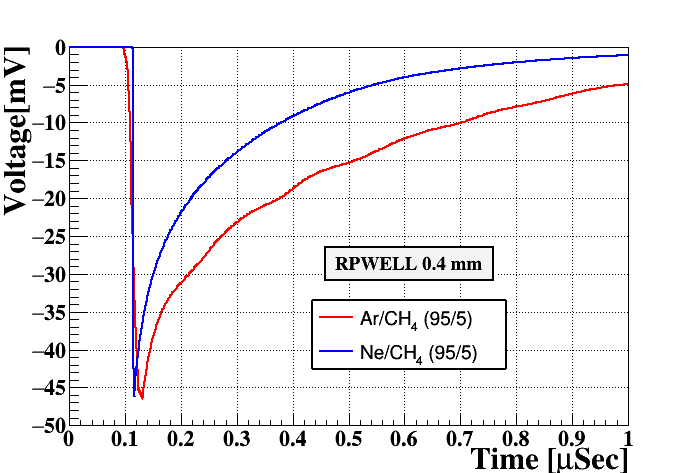}}
\caption{A comparison of the fast current signals in a $0.4~\mathrm{mm}$ thick RPWELL detector in $\mathrm{Ar}/5\%\mathrm{CH_{4}}$ and $\mathrm{Ne}/5\%\mathrm{CH_{4}}$: (a) measurement, (b ) simulation.}
\label{GasCurrent}
\end{figure}

The measured and simulated fast-current signals induced by x-rays in 
THGEM detector, recorded with the custom-made current-sensitive pre-amplifier 
are shown in Figure \ref{THGEMCurrent}. 
Due to the pre-amplifier shaping time, the THGEM pulses have a rise-time 
of $\sim25~\mathrm{nsec}$ and a width of $\sim100~\mathrm{nsec}$ 
(electron drift time within the induction gap). 
A comparison of measured and simulated fast current-signal shapes, recorded 
with the custom-made current preamplifier, in $0.8~\mathrm{mm}$ thick THWELL 
and RPWELL detectors are shown in Figure \ref{DetectorCurrent}. 
The THWELL and RPWELL detectors pulses are narrower; they have a faster 
rise-time, of $\sim10~\mathrm{nsec}$. 
The current peak is followed by a long ion tail, more pronounced in the 
simulation than in the measurement.
The electrode thickness does not affect the signals' rise time 
of RPWELL detectors as shown in Figure \ref{RPWELLCurrent}. 
The effect of the gas mixture on the signal shape in the RPWELL detector 
is shown in Figure \ref{GasCurrent}.

\subsection{Response to Cosmic Muons: Measurements and Simulations}

\begin{figure}[h]
\centering
\subfigure[]
{\label{Cosmic-1}\includegraphics[height=4cm]{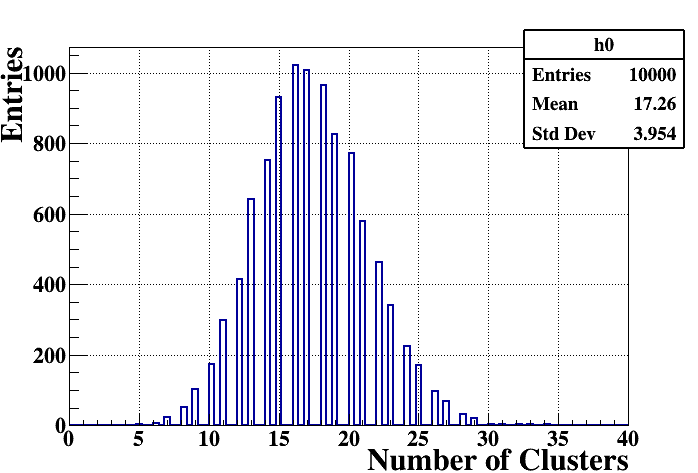}} \\
\subfigure[]
{\label{Cosmic-2}\includegraphics[height=4cm]{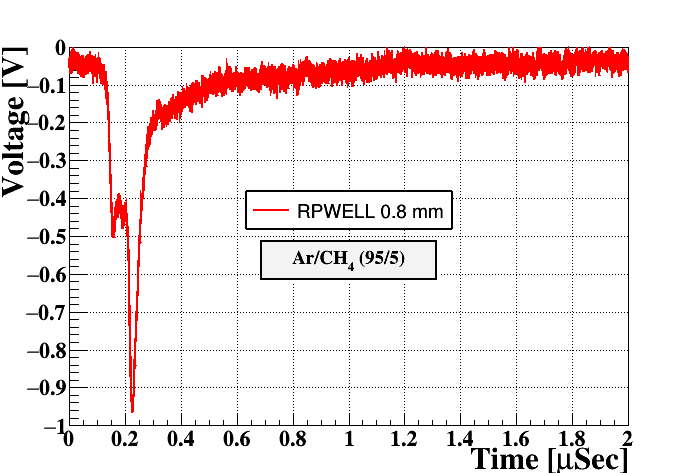}}
\subfigure[]
{\label{Cosmic-3}\includegraphics[height=4cm]{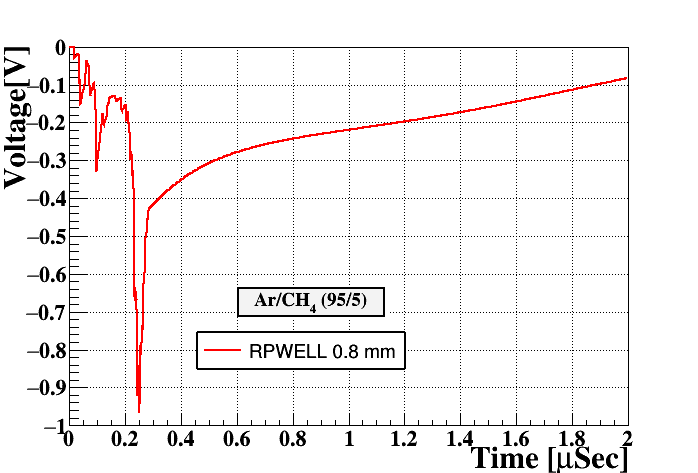}}
\caption{RPWELL response to cosmic muons interacting in a normal direction to the detector surface, in $\mathrm{Ar}/5\%\mathrm{CH_4}$ . (a) A simulated Muon-induced Number-of-clusters distribution; (b) measured signal from a fast pre-amplifier; (c) simulated fast pre-amplifier signal.}
\label{Cosmic}
\end{figure}

While the soft x-ray photons induce local electron clusters in each interaction 
in the $5~\mathrm{mm}$ drift gap, cosmic muons induce extended ionization 
charges along their tracks; a computed distribution of ionization clusters 
within the $5~\mathrm{mm}$ gap is shown in Figure \ref{Cosmic-1}. 
The electron drift time from a cluster positioned close to a hole could be 
at the order of a few $\mathrm{nsec}$; others, drifting to the hole from
distant locations generate trains of signals during a few $100~\mathrm{nsec}$. 
Due to its long shaping time, a typical signal of a 
charge-sensitive pre-amplifier will consist of fast electron-component signals
on top of the long ion-component. 
Signals from a current-sensitive pre-amplifier exhibit cluster-induced 
peaks as shown in Figure \ref{Cosmic-2} (measured) and \ref{Cosmic-3} (simulated).

\subsection{Response to accelerated Muons: Measurements and Simulation}

\begin{figure}[hbt]
\centering
\subfigure[]
{\label{APV-1}\includegraphics[height=3.5cm]{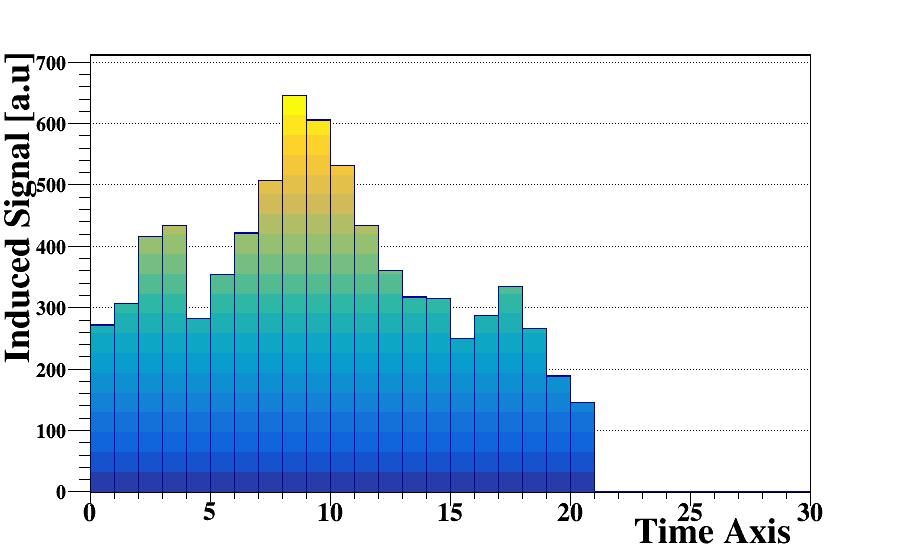}}
\subfigure[]
{\label{APV-2}\includegraphics[height=3.5cm]{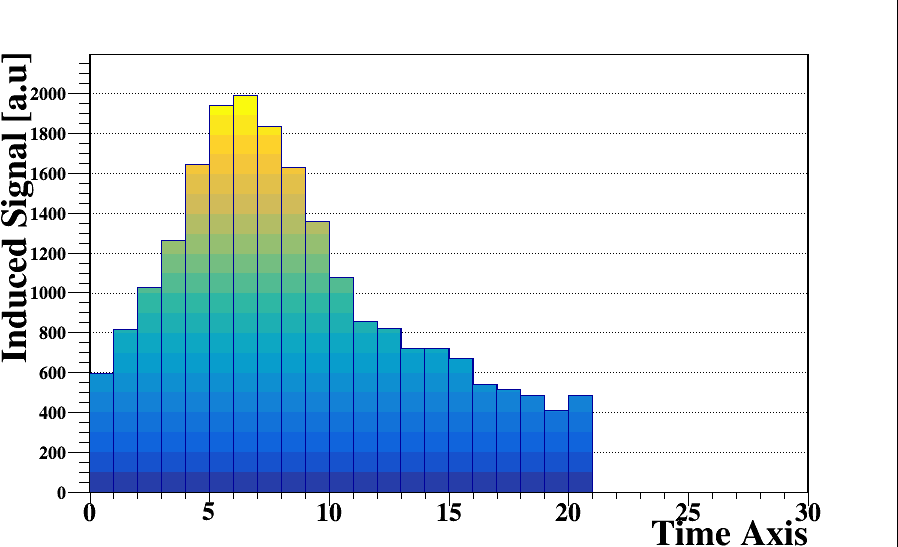}}
\subfigure[]
{\label{APV-3}\includegraphics[height=4cm]{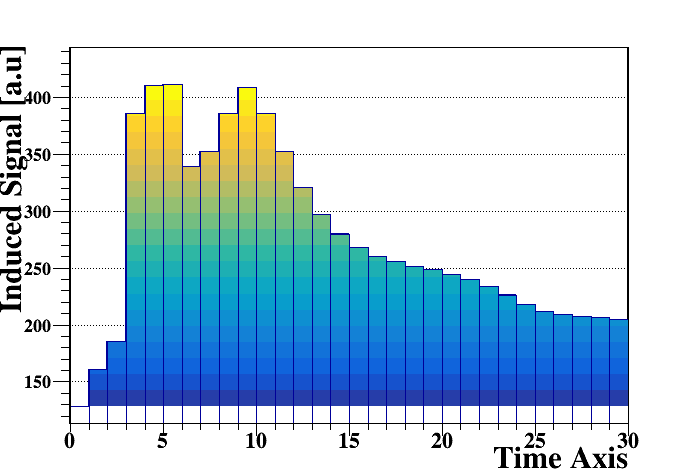}}
\subfigure[]
{\label{APV-4}\includegraphics[height=4cm]{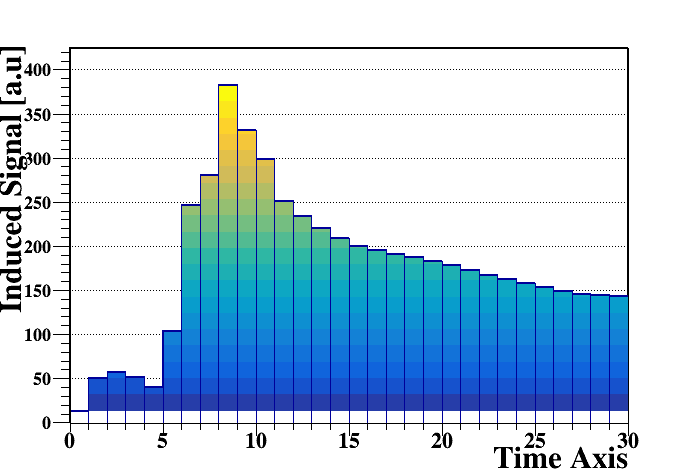}}
\caption{150 GeV muon-induced signals within a $5~\mathrm{mm}$ wide drift gap of an RPWELL detector operated in $\mathrm{Ne}/5\%\mathrm{CH_{4}}$: (a) and (b) 2 examples of measured signals from an APV25 chip; (c) and (d) 2 examples of simulated signals, taking into account the response function of the APV25 chip.}
\label{APV}
\end{figure}

We implemented our numerical model to simulate the signal shape measured with
the APV25 chip, having $\sim75~\mathrm{nsec}$ shaping time. 
Operated in ambient $\mathrm{Ne}/5\%\mathrm{CH_4}$ gas, the $0.8~\mathrm{mm}$ 
thick RPWELL detector (of $10~\times~10~\mathrm{cm^2}$) has been investigated 
with $150~\mathrm{GeV}$ muons (normal incidence) at CERN-SPS. 
The signal shapes varied from track to track, according to the number of 
primary clusters and their location within the $5~\mathrm{mm}$ long drift gap. 
As an example, two measured signals are shown in Figures \ref{APV-1} and \ref{APV-2}. 
In these measurements, we used a strip-patterned readout anode and the 
signal was induced on a few neighboring strips \cite{paper34}. 
For simplicity, in the simulation we assumed a continuous-non-patterned 
anode. A comparison of the shape of the signals induced on the central 
strip with the simulated one shows similar characteristics 
(Figure \ref{APV-3} and \ref{APV-4}).

\section{Discussion}

An evaluation of the signal formation in THGEM -like detectors is presented. 
The model-simulated signal shapes in THGEM (with induction gap), THWELL and
RPWELL were validated by experimental results . 
The simulations were performed combining Garfield and neBEM packages. 
The signal shapes have been simulated in $\mathrm{Ar}/5\%\mathrm{CH_4}$ 
and in $\mathrm{Ne}/5\%\mathrm{CH_4}$ under irradiation with 5.9 keV x -ray
photons, cosmic and relativistic muons. The raw signals were convoluted 
with the pre-amplifiers' response functions; the final 
signal forms were compared to that of measured signals in the three detector 
configurations investigated.

The soft-photon ``point ionization'' signals, measured by a fast preamplifier, 
differ from the ``extended ionization'' clustered muon-induced ones. 
The ``clustering'' effect is washed out using charge-sensitive pre-amplifiers of 
long shaping time; it is clearly observed with a fast current-pre-amplifier 
and with the APV25 chip having short shaping times. 
The difference between the signals in $\mathrm{Ar}$- and $\mathrm{Ne}$-based 
gas mixtures is reflected by the length of the ion-component tail (measured
with a charge preamplifier); it is shorter in the $\mathrm{Ne}$-based gas - 
due to higher ion mobility. 
The fast pre-amplifier, of $\sim50~\mathrm{nsec}$ shaping time, integrates 
only $5-10\%$ of the total charge. 
This effect should be taken into account when evaluating the detector gain. 
The variation of the detector thickness affects the slow ion's 
signal rise-time (with a charge-sensitive pre-amplifier), whereas the fast 
current (electrons mostly) signal does not reveal any significant change 
in the signal shape.

The effect of the RP is visible by comparing the ion component of the raw 
signal of the RPWELL to that of the THWELL. 
This affect is washed out when measuring with the charge sensitive 
pre-amplifier having long shaping time but is visible while using the 
fast pre-amplifier of short shaping time.

To summarise, rather good qualitative agreement was reached between the 
simulated and measured signal shapes, in all configurations investigated, 
under all irradiation conditions and gas mixtures. 
This confirms the success of the developed simulation tool - taking 
into account the important physics processes affecting the signal formation 
in these detectors as well as the response of the readout electronics. 
The model and the simulation kit provide a solid tool for
conducting similar studies in other gas-avalanche detector configurations.

\section{Acknowledgment}

We thank Dr. R. Veenhof and Prof. S. Mukhopadhyay for their valuable 
suggestions and comments. 
This research was supported in part by the I-CORE Program of the Planning 
and Budgeting Committee, the Nella and Leon Benoziyo Center for High 
Energy Physics at the Weizmann Institute of Science and by Grant No 712482 
from the Israeli Science Foundation (ISF).

\end{document}